# INFRARED SPECTROSCOPIC STUDY OF PHASE TRANSISIONS IN $A_1C_{60}$ COMPOUNDS (A=K,Rb,Cs)

DANIEL KOLLER, MICHAEL C. MARTIN, and LASZLO MIHALY
Department of Physics, SUNY at Stony Brook, NY 11794-3800

**Abstract** Optical measurements on doped $C_{60}$ films provide evidence for a transition from a high temperature conducting phase to a low temperature insulating phase in quenched $Rb_1C_{60}$ and $K_1C_{60}$ compounds, while $Cs_1C_{60}$ did not exhibit this behavior. For slow cooled samples our study confirms earlier results indicating phase separation of $K_1C_{60}$ to $K_3C_{60}$ and $C_{60}$, and the formation of a new $Rb_1C_{60}$ phase at low temperatures. Upon slow cooling $Cs_1C_{60}$ behaves similar to $Rb_1C_{60}$.

Molecular orbit considerations suggest that, as long as the unit cell contains one $C_{60}$ molecule, the alkaline metal doped $A_xC_{60}$ should be a metal for $0 < x < 6$. This conclusion may change dramatically if the electronic correlation effects are important, or if the unit cell becomes more complex. The investigation of the non-metallic $A_xC_{60}$ compounds is thus of great importane, since their behavior may shed light on the nature of electronic correlations in the $C_{60}$ compounds. The x=1 compound is the simplest and best candidate to begin this study.

Although evidence for a phase of lower than x=3 doping levels has been seen in early X-ray studies [1], the existence of $AC_{60}$ was first suggested in optical studies of $K_1C_{60}$ by Winter and Kuzmany [2]. They pointed out that the $K_1C_{60}$ phase decomposes to $K_3C_{60}$ and pure $C_{60}$ at temperatures below 80C. The first detailed X-ray investigation on $AC_{60}$ compounds was performed by Zhu and co-workers [3]. At high temperatures (above 450K) the materials were found to have a rocksalt structure, and the low temperature structure was approximately indexed as a rhombohedral structure. They found no evidence for phase separation. The NMR study of Tycko *et al.* [4] on $Rb_1C_{60}$ and $Cs_1C_{60}$ is in general agreement with these conclusions; for $K_1C_{60}$ the low temperature phase was seen to be phase separated. The different behavior between $K_1C_{60}$ on the one hand and $Rb_1C_{60}$ and $Cs_1C_{60}$ on the other hand was further confirmed by the photoemission study of Poirier *et al.* [5].

In a more recent investigation, Chauvet *et al.* [6] argue that the slow cooled, low temperature phase of $Rb_1C_{60}$ is orthorhombic below 350K. Similar to the earlier X-ray work, no phase separation was seen. Pekker *et al.* [7] points out that the structure may be produced by a Diels-Alder type polymerization of the $C_{60}$ molecules along one of the face diagonals of the fcc sub-lattice. Chauvet *et al.* [6] also established that the development of the orthorhombic phase can be prevented by cooling the sample at a fast rate. In these quenched samples the electron spin resonance spectrum indicates yet another phase transition, from a high temperature paramagnetic state to a low temperature non-magnetic state [6, 8]. The dramatic difference between the slow cooled and quenched samples have also been seen in IR spectroscopy by Martin *et al.* [9].

Here we report an optical transmission study of thin films with composition of primarily $A_1C_{60}$. The experiments have been performed by using a sample chamber developed for the infrared optical studies of doped $C_{60}$ films [10]. The permanently sealed sample cell allows for heating up to 300C, and exceptionally fast cooling, since the chamber can be immersed in liquid nitrogen. The IR radiation first passes through the silicon substrate wafer, then the doped film and then another wafer enclosing the sample cell. Transmission vs. frequency is calculated as a ratio between a spectrum obtained this way and the spectrum of the empty sample chamber, recorded at appropriate temperatures before sample deposition.

Here we concentrate on two features in the spectra: 1./ the IR active $F_{1u}(4)$ vibrational mode, and 2./the overall background transmission (represented by the transmission at 900 cm$^{-1}$, far away from the vibrational features). The $F_{1u}(4)$ mode can be conveniently used as an indicator of the electronic state of the molecule. As it is well established in other studies [11, 12], the frequency of the resonance shifts to lower values, and the line intensity gets stronger as the molecule is doped. To a smaller extent crystal fields (interaction between nearest neighbor $C_{60}$ molecules and alkaline ions) may also influence the mode frequency. In contrast to molecular vibrations, electronic transitions usually lead to broad features. In particular, excitations of the conduction electrons suppress the transmission from low frequencies up to the plasma frequency. The transmission at wavenumbers below the Drude relaxation rate is an indicator of the dc conductivity of the material.

In Fig. 1 we present spectra taken on samples exposed to slow cooling from 225C to −100C. The samples were first thermalized at high temperature for several hours, and then cooled slowly. The right hand side of the Figure emphasizes the behavior of the $F_{1u}(4)$ mode. The inspection of the 225C spectra reveals that the films are not entirely composed of the x=1 material, but other x values are also present [13]. Nevertheless, x=1 is the dominant phase (note that the $F_{1u}(4)$ line in the x=6 compound is very strong, so a small volume fraction of this compound leads to an observable resonance). Upon cooling, the $Cs_1C_{60}$ and $Rb_1C_{60}$ samples develop a characteristic splitting of the x=1 line. We associate this with the development of the new orthorhombic structure [6], in which $C_{60}$ molecules become very closely spaced in one direction, possibly forming a polymer chain [7]. On the other hand, in $K_1C_{60}$ the x=1 line broadens and then entirely disappears, while the intensity of the x=0 line (and, to a smaller extent, the x=3 line) increases. This observation, in accordance with the Raman study of Winter and Kuzmany [2], indicates phase separation.

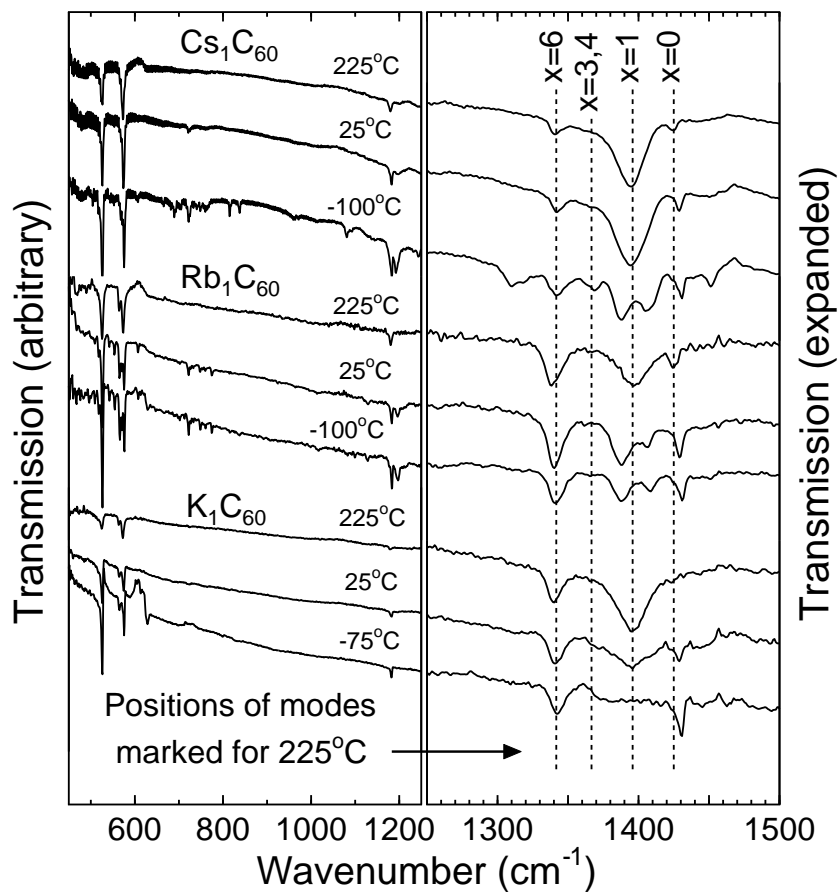

Figure 1. Transmission versus wavenumber for $AC_{60}$ samples subjected to slow cooling. The vertical scale is arbitrary. The frequency scale for the right panel is expanded. The dashed lines indicate the established positions of the $F_{1u}(4)$ mode for different values of x at 225C. For the x=0 phase, the temperature dependence of the line position is in accordance with earlier measurements [14]. The peaks seen in $Cs_1C_{60}$ around $1310 cm^{-1}$ and $1450 cm^{-1}$, as well as other weak features around $800 cm^{-1}$ probably correspond to vibrational modes which are IR forbidden by symmetry for a single $C_{60}$ molecule [15].

Subsequent heating of the samples shows that both the rocksalt to orthorhombic, and the phase separation transitions are reversible with a considerable hysteresis corresponding to a first order transition. Although not apparent from Fig. 1, the phase transitions are also seen in the average background transmissions of the samples; there is a few percent drop in the overall transmission as the samples are slowly cooled, and an increase upon warming.

As reported earlier, the quenched $Rb_1C_{60}$ samples exhibit a dramatic increase in the overall IR transmission [9]. This is best represented by plotting the trans-mission at a 900cm$^{-1}$ at various temperatures. In Fig. 2 we show the results for the three $AC_{60}$ compounds (for $K_1C_{60}$ and $Rb_1C_{60}$ the main features seen in the Figure were reproduced on another samples.) In these experiments, the samples were heated to 225C and kept at this temperature for about an hour. Then liquid $N_2$ was poured over the sample chamber, and the sample temperature was stabilized at around $-100$C. The temperature was elevated to a preset value, and spectra were taken. As the preset temperature approached the critical temperature of the first order structural/decomposition transition, the spectra became time dependent, corresponding to a strongly temperature dependent relaxation. For

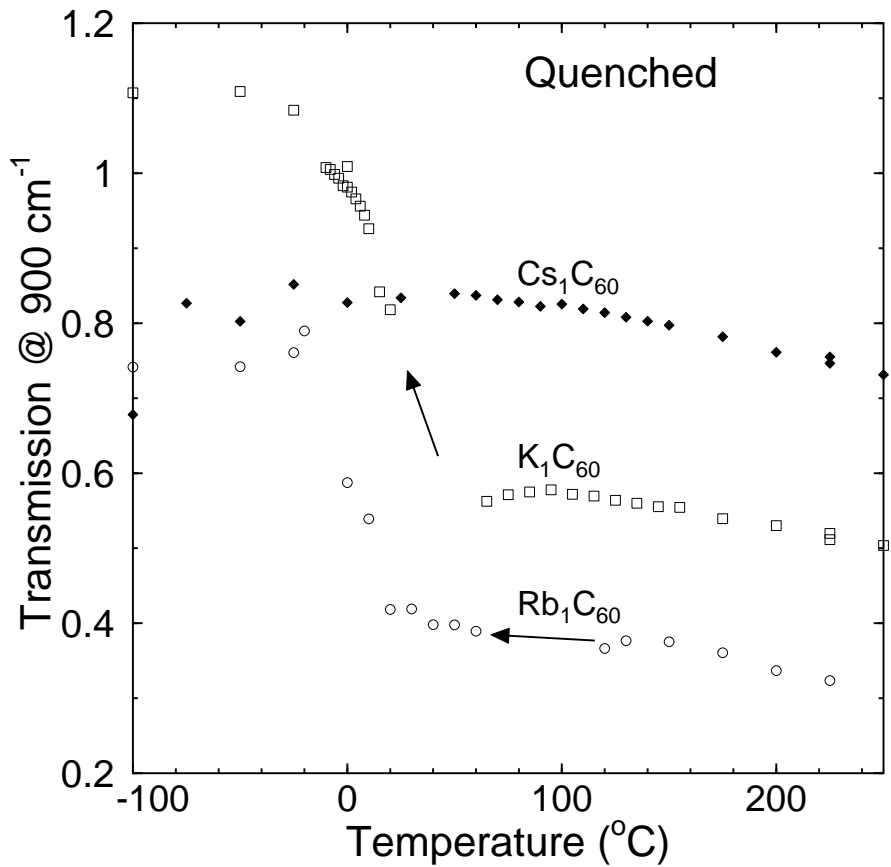

Figure 2. Transmission at 900cm$^{-1}$ versus temperature for the three samples in this study. The thicknesses of the films were 2$\mu$m for $Rb_1C_{60}$ and 1$\mu$m for $Cs_1C_{60}$ and $K_1C_{60}$. The arrows mark the temperature range where fast relaxation towards the thermodynamically stable structure prevented the study of the quenched phase. Transmission values of larger than unity are obtained as the $AC_{60}$ film reduces the amount of reflected light at the silicon/vacuum interface.

# INFRARED SPECTROSCOPIC STUDY OF $A_1C_{60}$

temperatures above the critical temperature, the spectra were taken in the equilibrium state. The slight downward curvature in the $K_1C_{60}$ data around 80C is due to the onset of the structural transition.

The spectra clearly show that in the *quench* experiments the $K_1C_{60}$ and $Rb_1C_{60}$ behave in a similar way, while $Cs_1C_{60}$ is different, since it does not show the sharp variation in the transmission. The transition from the high temperature/low transmission state to the low temperature/high transmission state is likely to be a second order phase transition, since it is occurs in fast cooled samples. Since the presence or absence of the transition correlates with the possibility of anion occupation at the tetrahedral sites (Cs is too large to fit in such sites), we speculate that the phase transition is somehow related to the ordering of alkaline metal ions in these sites. In principle, the transition may be the property of the quenched x=3 impurity phase in our samples. However, we quenched films of predominantly x=3 composition and no similar behavior was observed. Yet another explanation, that the transition is driven by a Fermi surface instability (like a charge or spin density wave transition), would imply that the high temperature rocksalt structure is metallic, and does not offer an explanation for the different behavior of the $Cs_1C_{60}$ film.

Although the transmission of the $Cs_1C_{60}$ sample is rather high, its temperature dependence is similar to that of the high temperature/low transmission state of $K_1C_{60}$ and $Rb_1C_{60}$. Note that $Cs_1C_{60}$ sample is thinner than the $Rb_1C_{60}$ film, also leading to higher transmission. Therefore we believe that the quenched $Cs_1C_{60}$ is qualitatively similar to high temperature state of the $Rb_1C_{60}$ and $K_1C_{60}$ samples.

In conclusion, we found that upon slow cooling $K_1C_{60}$ samples phase separate into $K_3C_{60}$ and pure $C_{60}$, whereas the behavior of $Rb_1C_{60}$ and $Cs_1C_{60}$ is different. For these compounds, new lines develop in the IR spectrum, interpreted as an indicator of the rocksalt - orthorhombic phase transition seen in $Rb_1C_{60}$ by Chauvet *et al.* [6]. Upon fast cooling, the $Rb_1C_{60}$ and $K_1C_{60}$ samples behave similarly, in that they exhibit a transition from a high temperature, low transmission state to a low temperature, high transmission state. Correspondingly, the high temperature phase is a good conductor (low optical transmission) and the low temperature phase is an insulator (high transmission). In contrast, the quenched $Cs_1C_{60}$ sample displays no similar transition.

ACKNOWLEDGMENTS

We are indebted to P.W. Stephens, G. Faigel, M. Tegze, A. Jánossy and L. Forro for valuable discussions. This work has been supported by NSF grant DMR9202528.